\documentclass[journal]{IEEEtran}
\ifCLASSINFOpdf
  \usepackage[pdftex]{graphicx}
  \graphicspath{{figures/}}
\else
  \usepackage[dvips]{graphicx}
  \graphicspath{{figures/}}
\fi
\usepackage{amsmath}
\usepackage{textgreek}

\usepackage{array}

\usepackage{multirow}

\usepackage{xcolor}

\newcommand{\tmu}{\textmu{}}
\newcommand{\talpha}{\mathrm{\alpha}}
\newcommand{\tbeta}{\mathrm{\beta}}
\newcommand{\tgamma}{\mathrm{\gamma}}
\newcommand{\teta}{\mathrm{\eta}}

\newcommand{\red}[1]{{\color{black}#1}}

\begin{document}
%
\title{Design of Compact and Efficient Silicon Photonic Micro Antennas with Perfectly Vertical Emission}
%
%
%

\author{Daniele~Melati,
        Mohsen~{Kamandar Dezfouli},
        Yuri~Grinberg,
        Jens~H.~Schmid,
        Ross~Cheriton,
        Siegfried~Janz,
        Pavel~Cheben,
        and~Dan-Xia~Xu
\thanks{Authors are with National Research Council Canada, Ottawa, ON K1A 0R6, Canada.  e-mail: daniele.melati@nrc-cnrc.gc.ca.}
\thanks{Manuscript received XXXXXX; revised XXXXXXX}}

%
%

\markboth{Journal of Selected Topics in Quantum Electronics,~Vol.~XX, No.~X, XXXXX~2020}%
{}
%


\IEEEspecialpapernotice{(Invited Paper)}

\maketitle

\begin{abstract}
Compact and efficient optical antennas are fundamental components for many applications, including high-density fiber-chip coupling and optical phased arrays. Here we present the design of grating-based micro-antennas with perfectly vertical emission in the 300-nm silicon-on-insulator platform. We leverage a methodology combining adjoint optimization and machine learning dimensionality reduction to efficiently map the multiparameter design space of the antennas, analyse a large number of relevant performance metrics, carry out the required multi-objective optimization, and discover high performance designs. Using a one-step apodized grating we achieve a vertical upward diffraction efficiency of almost 92\% with a 3.6 \textmu{}m-long antenna. When coupled with an ultra-high numerical aperture fiber, the antenna exhibits a coupling efficiency of more than 81\% (-0.9 dB) and a 1-dB bandwidth of almost 158 nm. The reflection generated by the perfectly vertical antenna is smaller than -20 dB on a 200-nm bandwidth centered at $\lambda$ = 1550 nm.
\end{abstract}


%
\IEEEpeerreviewmaketitle

\section{Introduction}
\begin{figure}[!t]
\centering
\includegraphics[width=0.9\columnwidth]{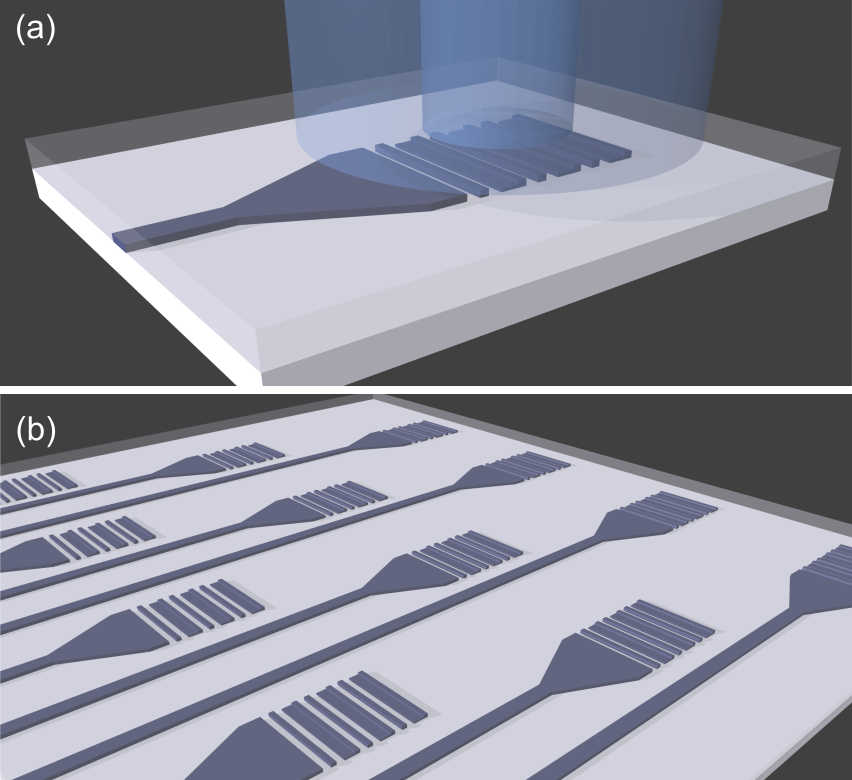}
\caption{Schematic of a grating-based micro-antenna used (a) as fiber coupler and (b) as a part of a dense integrated optical phased array.}
\label{fig_model}
\end{figure}

%
%
%
%
\IEEEPARstart{L}{ight} coupling between integrated photonic devices and the off-chip environment has always posed challenging research problems, especially for high-index-contrast platforms such as Silicon-On-Insulator (SOI). Antennas based on surface gratings have been widely used to interface integrated circuits with optical fibers (Fig. \ref{fig_model}(a)) as well as for free-space coupling applications such as integrated optical phased arrays (Fig. \ref{fig_model}(b)) \cite{taillaert_compact_2004,sun_large-scale_2013,meynard_sin_2020}. In both cases, antennas can be flexibly arranged in any desired pattern on the chip. A number of solutions have been proposed to improve their efficiency and directionality in diffracting light, including embedding Bragg or metal mirrors at the chip backside \cite{baudot_low_2014,benedikovic_subwavelength_2015}, use of silicon overlays \cite{tong_efficient_2018, vermeulen_high-efficiency_2010}, and multiple etch steps \cite{taillaert_compact_2004, vermeulen_high-efficiency_2010, roelkens_high_2007, alonso-ramos_fiber-chip_2014, bozzola_optimising_2015}. However, other limitations still exist. Dimensions of SOI surface gratings are typically on the order of few tens of microns and a significant size reduction is not easy to achieve without compromising efficiency \cite{pita_design_2018}. This can be a limiting factor for applications requiring a high integration density, e.g. optical phased arrays or multi-device photonic circuits with numerous fiber interfaces \cite{sun_large-scale_2013, pita_design_2018, fatemi_nonuniform_2019}. Moreover, the diffraction angle of surface gratings normally depends rather strongly on the wavelength of the light, limiting their operational bandwidth compared to edge fiber couplers or end-fire antennas \cite{benedikovic_sub-decibel_2019,guan_hybrid_2015}. The diffraction angle is often designed to be slightly offset from the chip surface normal direction to avoid high reflections into the input waveguide from the second-order diffraction \cite{watanabe_perpendicular_2017}. This can introduce additional challenges in fiber coupling or alignment with other antennas and increase packaging complexity.

Depending on the chosen structure, the behaviour of surface gratings can be controlled by a large number of design parameters. For this reason, grating design has benefited from the use of optimization techniques such as the genetic or particle swarm algorithms \cite{tong_efficient_2018, watanabe_perpendicular_2017}. More recently, inverse design using gradient-based local optimization algorithms in combination with adjoint simulations \cite{borel_topology_2004} has been demonstrated to be particularly efficient in exploring the large design spaces generated by complex grating structures \cite{michaels_inverse_2018, su_fully-automated_2018}.
Although extremely powerful, both global and local optimization strategies usually generate only a single or a handful of designs optimized for the selected objective and do not shed light on the characteristics of the design space and the influence of the design parameters on the behaviour of complex devices. \red{Moreover, the simultaneous optimization of multiple performance metrics is often non-trivial and computationally intensive. A commonly exploited approach is to carefully craft an objective function that properly weighs several competing terms to obtain the desired result \cite{michaels_hierarchical_2020}. Other solutions look for Pareto fronts to solve optimization problems involving multiple objectives, i.e. sets of solutions for which improving a given objective necessarily deteriorates at least another one \cite{gagnon_multiobjective_2013}.} In practical implementations, often a number of performance metrics has to be taken into account. High diffraction efficiency and directionality, low reflection, compact footprint, and fabrication tolerance are some desirable features common to both fiber and free-space coupling applications. As fiber couplers in high-data-throughput interfaces, also high modal overlap with the fiber (i.e. low interface losses) and large bandwidth are fundamental performance metrics of the antennas. An efficient, multi-objective, and flexible optimization strategy capable of balancing all these aspects is hence a valuable tool.

\begin{figure}[!t]
\centering
\includegraphics[width=0.9\columnwidth]{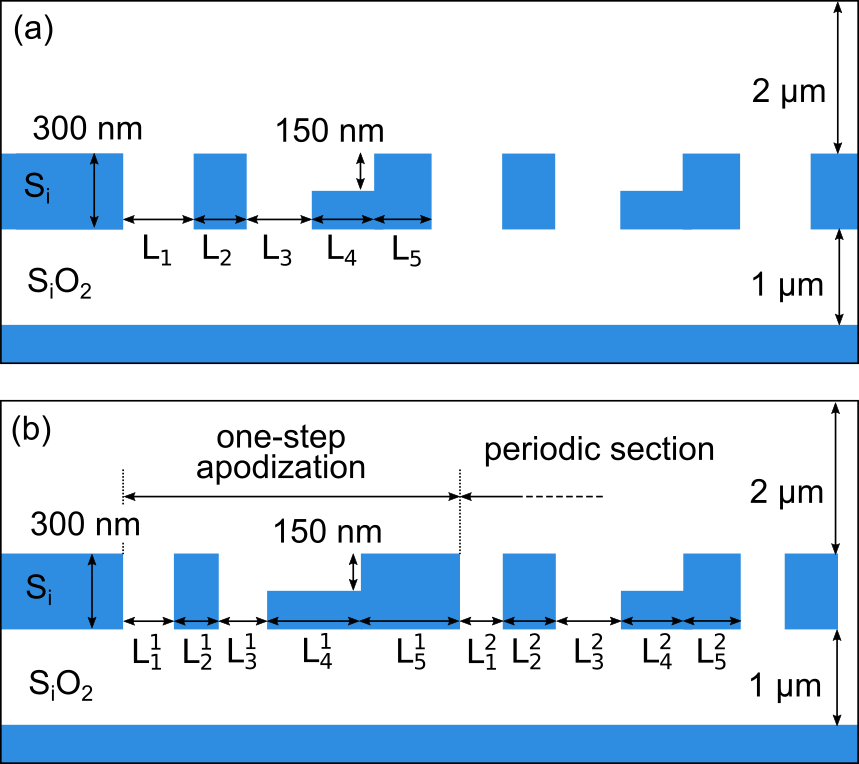}
\caption{2D cross-section of the grating for (a) the periodic design and (b) the one-step apodized design. Segment lengths are the design parameters defining a 5-dimensional (for a fully periodic design) or 10-dimensional (for a one-step apodized design) parameter space.}
\label{fig_struct}
\end{figure}
In this paper, we present the design of compact, highly efficient, and ultra-broadband micro-antennas based on surface gratings. These antennas have a perfectly vertical emission, i.e. the beam centre axis is aligned along the direction normal to the chip surface, while maintaining low reflections over a large spectral range. An approach we recently proposed \cite{melati_mapping_2019, melati_design_2020} combining inverse design based on adjoint simulations \cite{lalau-keraly_adjoint_2013} and machine-learning dimensionality reduction is used for the design. This methodology allows to effectively map and characterize the multiparameter design space of the antennas. As a result, we are able to obtain a wealth of information on the behaviour of a range of antenna designs, making possible the analysis of a number of relevant performance metrics and their mutual trade-offs. By exploiting a 300-nm-thick silicon core and a structure made of a subwalength pillar and a partially etched L-shaped segment \cite{watanabe_perpendicular_2017, melati_mapping_2019} we achieve a vertical upward diffraction efficiency of almost 92\% at $\lambda$ = 1550 nm with an antenna of only 3.6 \textmu{}m in size. When coupled to an ultra-high numerical aperture fiber with mode field diameter of 3.2 \textmu{}m, the antenna has a maximum coupling efficiency of more than 81\%, or -0.9 dB, with a 1-dB bandwidth of almost 158 nm. Reflection is lower than -20 dB over a 200 nm bandwidth between $\lambda$ = 1450 nm and $\lambda$ = 1650 nm. Fabrication tolerance is also analysed, predicting a fabrication yield of 88\% for the selected design under common process variability.

\begin{figure}[!t]
\centering
\includegraphics[width=0.9\columnwidth]{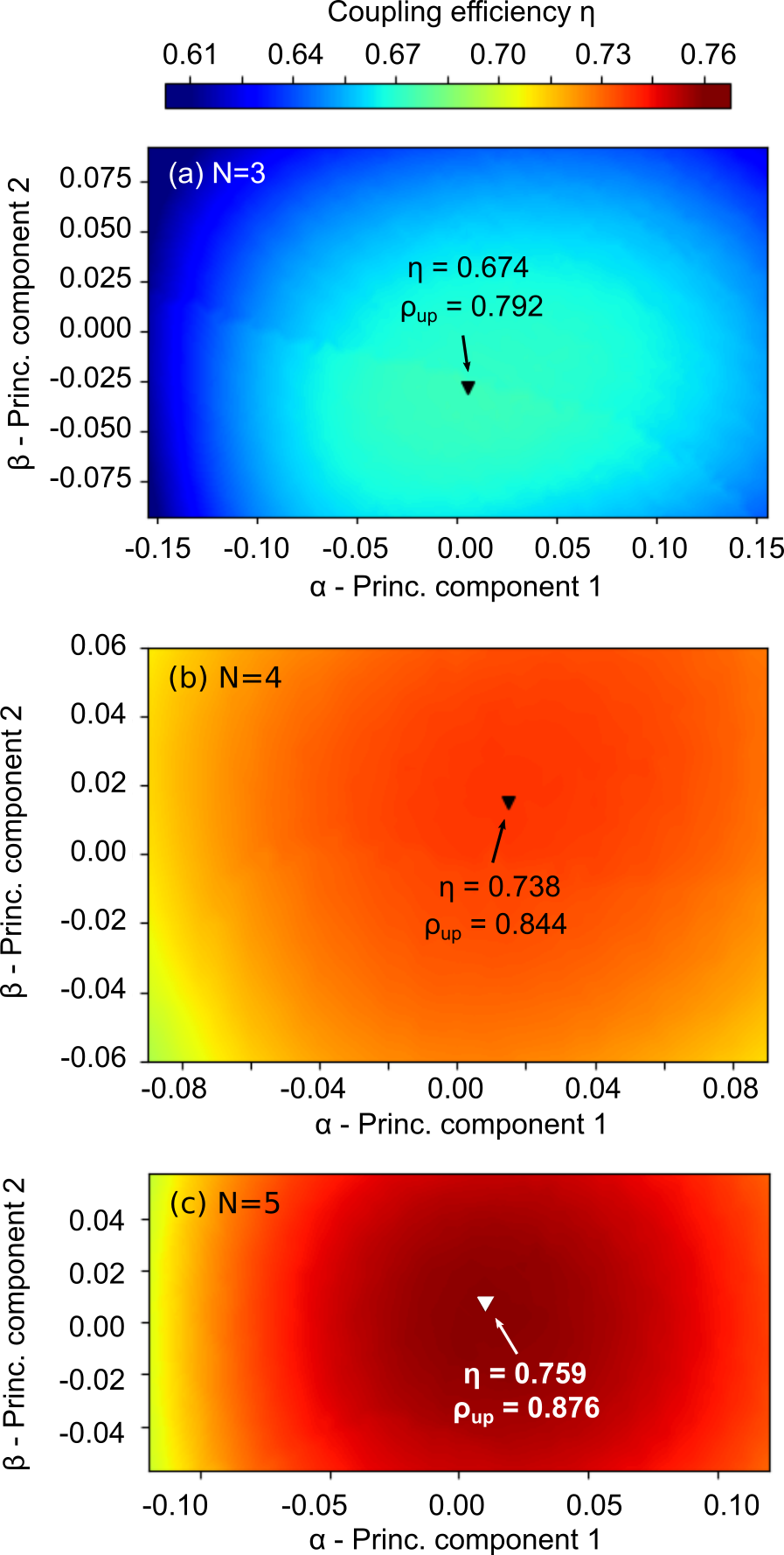}
\caption{Fiber coupling efficiency $\teta$ of antennas based on periodic gratings at $\lambda$ = 1550 nm as a function of parameters $\talpha$ and $\tbeta$ for a design with (a) 3 periods, (b) 4 periods, and (c) 5 periods. The global optimum in each case is indicated with a triangle and the corresponding upward diffraction efficiency $\rho_{up}$ and fiber coupling efficiency $\teta$ are reported.}
\label{fig_ce_periodic}
\end{figure}
The paper is organized as follow. Section \ref{sec_periodic} summarizes the design methodology and investigates the performances that can be achieved with micro-antennas based on periodic grating structures. The analysis of one-step apodized gratings is presented in Sec. \ref{sec_apodized} where several performance metrics and achievable trade-offs are discussed. Based on this analysis, Sec. \ref{sec_optimization} reports on the multi-objective optimization of the apodized design. The impact of stochastic fabrication variability is discussed in Sec. \ref{sec_tolerance}, including an estimation of the expected device fabrication yield. Conclusions are drawn in Sec. \ref{sec_conclusions}. Finally, Appendix \ref{app_1} includes additional details on the design methodology and Appendix \ref{app_2} discusses some physical insights on the grating behavior. 

\section{Periodic grating design}
\label{sec_periodic}
\begin{figure}[t]
\centering
\includegraphics[width=0.9\columnwidth]{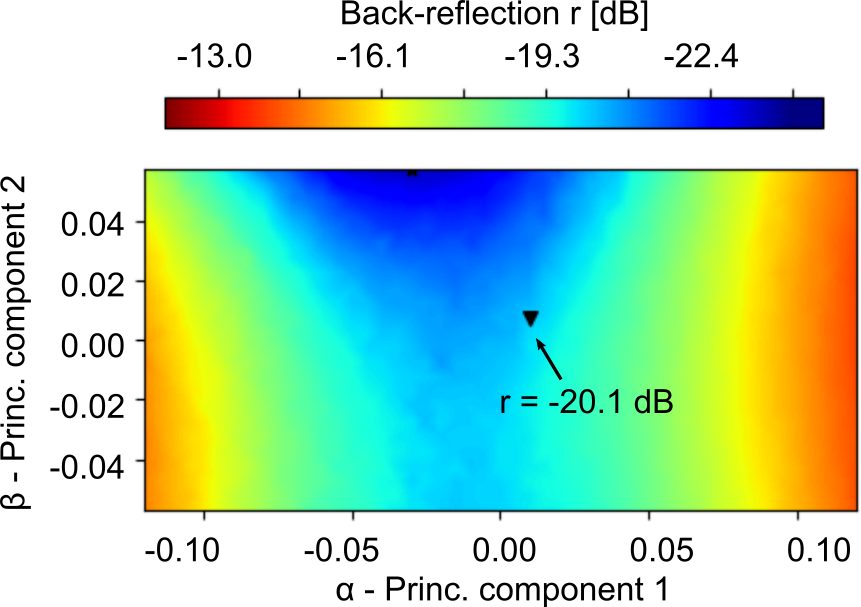}
\caption{Simulated back-reflection at $\lambda$ = 1550 nm for an antenna based on a periodic grating with 5 periods as a function of $\talpha$ and $\tbeta$. The triangular mark is the design with highest coupling efficiency as shown in Fig. \ref{fig_ce_periodic}(c)}
\label{fig_br_periodic}
\end{figure}
The antenna structure considered in this work is based on a surface grating as schematically shown in Fig. \ref{fig_struct}. Each period of the grating consists of a 300-nm-thick pillar and an L-shaped section with a partial etch of 150 nm. The L-shape provides blazing to increase the fraction of power diffracted upwards and improve the grating directionality \cite{benedikovic_l-shaped_2017} while the pillar reduces back-reflection by destructive interference \cite{watanabe_perpendicular_2017, melati_mapping_2019}. The use of a thicker silicon layer compared to the standard 220 nm increases the grating scattering strength and hence reduces the required number of periods (i.e., the antenna footprint) to achieve a target efficiency \cite{bozzola_optimising_2015,xu_silicon_2014}.
A silica upper cladding of 2 \textmu{}m in thickness and 1 \textmu{}m of buried silica oxide are assumed. 
For the initial periodic structure shown in Fig. \ref{fig_struct}(a), the $k$-th device in the design space is represented by five parameters ${\mathbf{L}_k}=[L_{1,k}, L_{2,k}, L_{3,k}, L_{4,k}, L_{5,k}]$, i.e., the lengths of each segment within a grating period. The design space is explored using the machine learning design technique described in \cite{melati_mapping_2019, melati_design_2020}. To ensure a perfectly vertical emission, we assumed a fiber to be placed vertically on top of the antenna with the fiber facet in direct contact with the top of the upper cladding and we collect a sparse set of designs with high fiber coupling efficiency. This is obtained by multiple runs of a local optimizer based on box-constrained limited-memory Broyden-Fletcher-Goldfarb-Shanno algorithm (L-BFGS-B) and adjoint simulations \cite{lalau-keraly_adjoint_2013}. We then apply linear principal component analysis (PCA) to these good designs. Similar to that reported in our previous work \cite{melati_mapping_2019}, PCA reveals that two principal components are sufficient to accurately represent the entire pool of good designs, i.e.
\begin{equation}
    \label{eq_pca}
        \mathbf{L}_k\simeq \alpha_k\mathbf{V}_1+\beta_k\mathbf{V}_2+\mathbf{C},
\end{equation}
where $\mathbf{V}_1$ and $\mathbf{V}_2$ are the two principal components and $\mathbf{C}$ is a constant vector. Instead of five lengths, we can hence represent a good antenna design (i.e., an antenna with high coupling efficiency) by using only the two parameters $\alpha_k$ and $\beta_k$. Since all good designs lie on the 2D $\talpha$-$\tbeta$ sub-space, the rest of the design space can be excluded from further investigation. The $\talpha$-$\tbeta$ sub-space can be rapidly and exhaustively mapped by parameter sweeps computing not only the coupling efficiency but also any other required performance, allowing a comprehensive analysis of the device behavior. Further details on the design methodology are reported in Appendix \ref{app_1}.

Antenna simulations are performed using the commercial 2D-FDTD solver from Lumerical. The light is launched into the antenna by the fundamental TE mode of the input waveguide at the left of Fig. \ref{fig_struct}(a). Silicon and silica refractive indices are 3.478 and 1.448 at $\lambda$ = 1550 nm, respectively, and dispersion is also taken into account \cite{xu_empirical_2019}. The silicon substrate is included in the simulation even though its effect is normally negligible due to the high directionality of the grating originated from its vertical asymmetry. The mode of the fiber is modeled with a Gaussian function with a mode field diameter of 3.2 \textmu{}m at $\lambda$ = 1550 nm.  The longitudinal position of the fiber along the antenna is optimized for each simulation in order to maximize the fiber coupling efficiency, calculated as:
\begin{equation}
    \label{eq_ce}
        \teta=\rho_u \cdot \varphi,
\end{equation}
where $\rho_u$ is the fraction of the injected optical power diffracted upwards and $\varphi$ is the overlap integral between the diffracted field and the Gaussian function.

\begin{table}[t]
\renewcommand{\arraystretch}{1.3}
\caption{Structural parameters of selected antenna designs based on periodic and one-step apodized gratings.}
\label{table_designs}
\centering
\begin{tabular}{|c|c|>{\centering\arraybackslash}m{2.6cm}|c|}
    \hline
     \multicolumn{2}{|c|}{} & $\mathbf{L}$ [nm] & Total length [\tmu{}m]\\
    \hline
    \multirow{3}{*}{Periodic} & N = 3 & [117 43 285 161 159] & 2.30\\
                              & N = 4 & [116 50 217 177 161] & 2.88\\
                              & N = 5 & [112 54 188 176 171] & 3.51 \\
    \hline
    \multirow{2}{*}{Apodized} & A & [67 70 120 136 224 145 41 235 155 171] & 3.60\\
                              & B & [57 93 118 90 277 143 41 225 156 171] & 3.58 \\
    \hline
\end{tabular}
\end{table}
\begin{figure*}[!ht]
\centering
\includegraphics[width=0.95\textwidth]{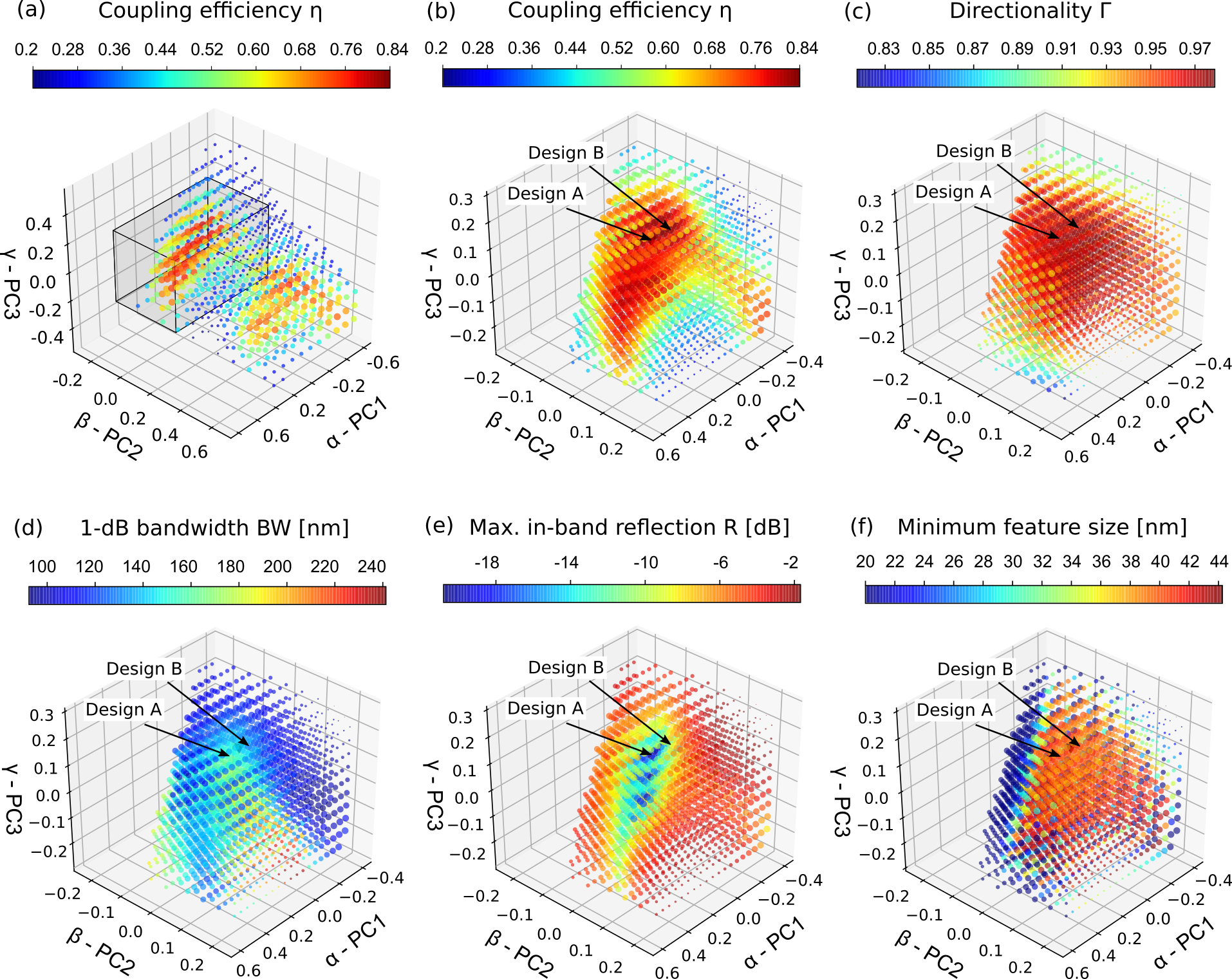}
\caption{Analysis of multiple performance metrics for the one-step apodized designs. (a) Fiber coupling efficiency on a wide region of the $\talpha$-$\tbeta$-$\tgamma$ sub-space. The gray box highlights the smaller region including the highest achievable $\teta$ and explored in the following panels. (b) Coupling efficiency in the region of interest; (c) grating directionality; (d) 1-dB bandwidth around the wavelength of the peak coupling efficiency; (e) maximum reflection generated by the grating in the wavelength range from 1450 nm to 1650 nm; (f) minimum feature size. Design A is selected based on the criteria described in Sec. \ref{sec_optimization} while deisgn B has the largest possible coupling efficiency $\teta$.}
\label{fig_apod_design}
\end{figure*}
We consider here three different periodic grating designs with N = 3, N = 4, and N = 5 periods, respectively, to find the minimum acceptable antenna length that does not compromise diffraction efficiency. In the three cases, we sweep $\talpha$ and $\tbeta$ to map coupling efficiency and reliably identify the global optimum (see Appendix \ref{app_1} for details). The results for $\lambda$ = 1550 nm are shown in Fig. \ref{fig_ce_periodic}(a)-(c) where the best design (highest $\teta$) is also marked. A larger number of periods allows to increase the highest achievable fiber coupling efficiency $\teta$ from 0.674 (-1.71 dB) to 0.759 (-1.20 dB) and the corresponding upward diffraction efficiency $\rho_u$ from 0.792 to 0.876. Using 6 periods does not significantly improve efficiencies any further and results are not shown. For all the three cases, the L-shaped structure in the considered grating geometry guarantees a high directionality, defined as:
\begin{equation}
    \label{eq_directionality}
        \Gamma = \frac{\rho_u}{\rho_u+\rho_d},
\end{equation}
with $\rho_u$ and $\rho_d$ the fraction of the injected optical power diffracted upwards and downwards, respectively. For N = 3, $\Gamma$ = 0.980, for N = 4, $\Gamma$ = 0.939, and with N = 5, $\Gamma$ = 0.947. Structural parameters of the design with highest $\teta$ are reported in Tab. \ref{table_designs} for all the three cases. It is interesting to notice how minimum feature size grows from 43 nm to 54 nm when N is increased from 3 to 5. These small features are not surprising giving the compactness of the designs and can be achieved by advanced 193 nm immersion deep ultraviolet lithography \cite{thomson_roadmap_2016, selvaraja_193nm_2014}. Larger feature sizes can be obtained without compromising efficiency exploiting a subwavelength metamaterial \cite{cheben_subwavelength_2018} in the direction transverse to propagation, as described in \cite{kamandar_perfectly_2020}.

For the five-period design (N = 5) the 1-dB bandwidth is about 133 nm. For this antenna, the simulated reflection (i.e., the fraction of power coupled to the counter-propagating fundamental TE mode of the input waveguide) as a function of $\talpha$ and $\tbeta$ is reported in Fig. \ref{fig_br_periodic} for $\lambda$ = 1550 nm. As can be seen, for the design with the highest fiber coupling efficiency (triangular mark) back-reflection is lower than -20 dB. Smaller back-reflections up to -23 dB can be achieved at the cost of a slightly lower fiber coupling efficiency of 0.747. The grating with N = 5 combines high diffraction and fiber coupling efficiency, small back-reflection, and a compact total length of only 3.51 \tmu{}m. Further analyses on how different parameters influence the behaviour of this design are reported in Appendix \ref{app_2}.

\section{One-step apodization}
\label{sec_apodized}
In a periodic grating, the overlap integral $\varphi$ in Eq. \ref{eq_ce} poses a theoretical limit of approximately 0.8 to the maximum achievable fiber coupling efficiency, due to the mismatch between the exponentially decaying intensity profile generated by the grating and the Gaussian-like shape of the fundamental mode of an optical fiber \cite{taillaert_compact_2004}. In order to increase coupling efficiency, the diffracted field profile can be shaped by apodizing the grating structure, consequently improving the overlap with the fiber mode \cite{xu_silicon_2014}. We adopt a one-step apodization strategy \cite{benedikovic_high-efficiency_2014}.

Based on the results shown in the previous section for the antennas based on periodic gratings, we design a grating structure where one period with a first set of structural parameters (one-step apodization) is followed by four identical periods (periodic section) with a second set of parameters. This structure is schematically shown in Fig. \ref{fig_struct}(b). The design space is hence 10 dimensional in this case and a given antenna design $k$ is represented by the parameters ${\mathbf{L}_k}=[L_{1,k}^1, \cdots, L_{5,k}^1, L_{1,k}^2,\cdots, L_{5,k}^2]$. The design methodology is identical to the one described in Sec. \ref{sec_periodic} with the difference that now the sub-space of good designs discovered through PCA grows from 2D to 3D, that is:
\begin{equation}
    \label{eq_pca_apod}
        \mathbf{L}_k\simeq \talpha_k\mathbf{V}_1+\tbeta_k\mathbf{V}_2+\tgamma_k\mathbf{V}_3+\mathbf{C},
\end{equation}
where $\mathbf{V}_1$, $\mathbf{V}_2$, and $\mathbf{V}_3$ are the three principal components and $\mathbf{C}$ is a constant vector. The principal components are $\mathbf{V}_1$ = [0.139, -0.313, 0.025, 0.607, -0.705, 0.022, -0.001, 0.129, -0.020, -0.001] \tmu{}m, $\mathbf{V}_2$ = [-0.125, 0.890, 0.221, 0.348, -0.106, 0.075 , -0.007 ,  0.015, -0.066,  0.032] \tmu{}m, and $\mathbf{V}_3$ = [-0.181, -0.300, 0.790, 0.258, 0.325, -0.014, -0.035, -0.146, -0.170, 0.174] \tmu{}m. Instead of 10 segment lengths, parameters $\talpha$, $\tbeta$, and $\tgamma$ are used to represent one-step apodized devices.

\begin{figure}[t]
\centering
\includegraphics[width=0.9\columnwidth]{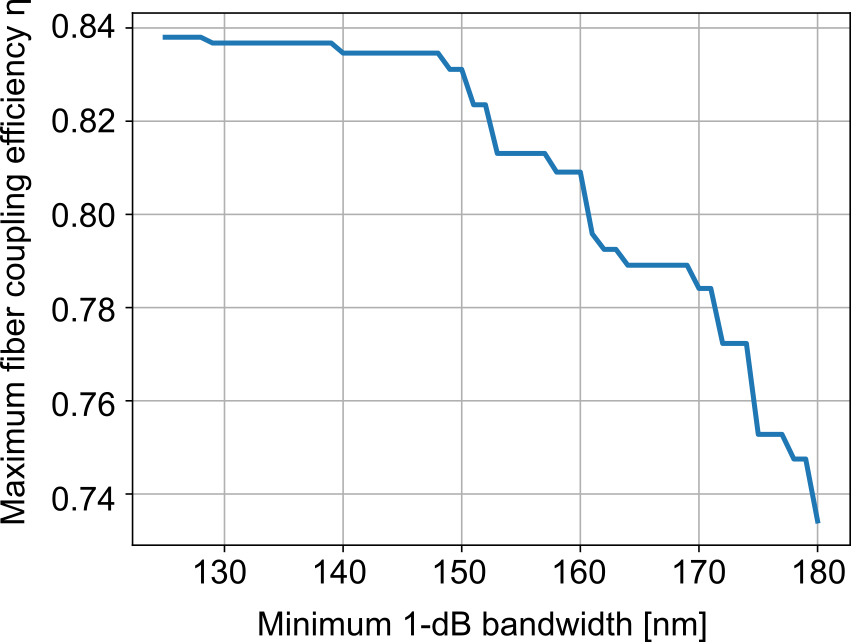}
\caption{Pareto frontier for the maximum achievable fiber coupling efficiency with the one-step apodized grating as a function of the corresponding 1-dB bandwidth.}
\label{fig_apod_bece_opt}
\end{figure} 
A number of performance metrics are calculated as a function of these three parameters, providing a wealth of information to support design selection. Figure \ref{fig_apod_design} reports on the mapping results. To ease visualization, the size of the dots in all sub-panels are proportional to the fiber coupling efficiency of the corresponding design. A first low-resolution mapping of $\teta$ is performed to explore a wide portion of the $\talpha$-$\tbeta$-$\tgamma$ sub-space, as shown in Fig. \ref{fig_apod_design}(a). Two potentially interesting areas with good fiber coupling efficiency designs (at $\lambda$ = 1550 nm) can be identified. The area toward lower values of $\tbeta$ hosts the highest efficiencies and is selected for further analysis. A zoom-in of $\teta$ in the region highlighted by the gray box is shown in Fig. \ref{fig_apod_design}(b). A broad continuous area with $\teta>0.75$ is now clearly visible, with a maximum fiber coupling efficiency of 0.838 (-0.77 dB), i.e., an improvement of about 10\% compared to the best periodic design with 5 periods shown in Sec. \ref{sec_periodic}. For the same region, Fig. \ref{fig_apod_design}(c) shows the grating directionality at $\lambda$ = 1550 nm as a function of $\talpha$, $\tbeta$, and $\tgamma$. Almost all the designs included in the mapped region show $\Gamma>0.9$ Among the designs with $\teta>0.75$, the highest achievable directionality is $\Gamma = 0.98$, with $\teta = 0.77$ (-1.14 dB).

The use of a strong and short grating also leads to an exceptionally wide operational bandwidth for the antenna, as shown in Fig. \ref{fig_apod_design}(d) where the 1-dB bandwidth centered around the peak coupling efficiency is reported. In this case designs with the widest possible bandwidth are not of interest because they show low fiber coupling efficiencies (see Fig.\ref{fig_apod_design}(b)). For designs with $\teta>0.75$, the highest achievable 1-dB bandwidth is BW = 178 nm between $\lambda$ = 1466 nm and $\lambda$ = 1644 nm with $\teta = 0.753$ at $\lambda$ = 1550 nm.
Figure \ref{fig_apod_design}(e) shows the maximum reflection R in the 200-nm wavelength range from $\lambda$ = 1450 nm to $\lambda$ = 1650 nm as a function of $\talpha$, $\tbeta$, and $\tgamma$. This is a more meaningful measure compared to the reflection generated at a single wavelength because it guarantees a consistent behaviour across the entire wide operational bandwidth of the antenna. An area of designs with low reflection can be identified overlapping the high-$\teta$ area. The lowest achievable in-band reflection is R = -20.8 dB, corresponding to a back-reflection at $\lambda$ = 1550 nm of -21 dB. Interestingly, the lowest achievable reflection at 1550 nm is -51.7 dB but this design has a behaviour much more sensitive to wavelength variations, with a maximum reflection in the 1450 nm - 1650 nm bandwidth as high as R = -13.5 dB. Lastly, Fig. \ref{fig_apod_design}(f) plots the minimum feature size. Dimensions below 20 nm are not relevant and are clipped to improve visual clarity. A large area with minimum feature size above 40 nm is identified, a range comparable with periodic designs.

\section{Multi-objective optimization}
\label{sec_optimization}
\begin{table}[t]
\renewcommand{\arraystretch}{1.3}
\caption{Performance of the apodized grating design selected according to Eq. \ref{eq_opt_problem} (A) and the design with the best coupling efficiency (B).}
\label{table_apod_performance}
\centering
\begin{tabular}{|c|c|c|c|c|c|c|}
    \hline
    & $\teta$ & $\rho_u$ & $\Gamma$ & R [dB] & BW [nm] & Y\\
    \hline
    A & 0.813 & 0.919 & 0.979 & -20.8 & 157.9 & 88\% \\
    B & 0.838 & 0.910 & 0.973 & -10.4 & 128.9 & 97\% \\
    \hline
\end{tabular}

\vspace{0.2cm}
$\teta$ fiber coupling efficiency; $\rho_u$ upward diffraction efficiency; $\Gamma$ directionality; R maximum reflection in the 1450 nm - 1650 nm wavelength range; BW 1-dB bandwidth; Y device yield.
\end{table}
Dealing with multi-objective optimization commonly requires designing for the optimizer an objective function that includes and properly weighs all the desired performance metrics. Often this complex process is based on trial and error and requires the designer's experience to find the correct balance between all the terms and guide the optimizer to the proper solution. The possibility to map the design space, as described in Sec. \ref{sec_apodized}, makes multi-objective optimization much easier, essentially reducing it to a post-processing look-up exercise.

From this point of view, it is of particular interest to investigate the relation between fiber coupling efficiency and 1-dB bandwidth, obtaining a form of Pareto frontier for an optimization problem involving these two quantities. This analysis can be easily done here by exploiting the data described above (only designs with minimum feature size above 40 nm are considered). Figure \ref{fig_apod_bece_opt} shows the maximum fiber coupling efficiency that can be achieved by the one-step apodized design requiring 1-dB bandwidth being above a threshold comprised between 125 nm and 180 nm. A clear trade-off is highlighted. A bandwidth threshold BW = 125 nm allows to reach the global maximum $\teta$ = 0.838 (see also Fig. \ref{fig_apod_design}(b)). If a wider bandwidth is desired, this has to necessarily come at the expense of a reduced efficiency. Requiring $\mathrm{BW} \geq 150$ nm, limits fiber coupling efficiency to $\teta \leq 0.831$. Further bandwidth increments quickly degrade fiber coupling efficiency, with $\teta \leq 0.734$ for $\mathrm{BW} \geq 180$ nm.

The availability of this thorough analysis allows the designer to easily select the best antenna design based on multiple objectives. In this case, we want an antenna with the highest possible fiber coupling efficiency, also ensuring a maximum reflection R in the 1450 nm - 1650 nm wavelength range below -20 dB, a 1-dB bandwidth BW larger than 150 nm, and minimum feature size above 40 nm. This corresponds to the optimization problem:
\begin{equation}
\label{eq_opt_problem}
\begin{aligned}
& \underset{L_1^1 \cdots L_5^2}{\text{maximize}} & & \teta(L_1^1 \cdots L_5^2) \\
& \text{subject to} & & \mathrm{R} \leq \text{-20 dB}\\
& & & \mathrm{BW} \geq \text{150 nm}\\
& & & L_i^m \geq \text{40 nm}, \; i = 1, \ldots, 5, \; m = 1,2.
\end{aligned}
\end{equation}
Solving Eq. \ref{eq_opt_problem} directly as a global optimization problem would pose significant challenges due to multiple constraints (or alternatively a complex objective function with multiple terms to be properly weighed) and the difficulty in navigating the 10D design space. Moreover, some of the constraints such as limits on bandwidth and reflection became clear only after the analysis presented in Figs. \ref{fig_apod_design} and \ref{fig_apod_bece_opt} that highlighted absolute achievable limits for these quantities. Lastly, constraints could change for different applications. With the information available here, a new multi-objective optimization simply requires another look-up operation in the different performance metrics while directly solving the optimization problem would require to start from scratch, without the possibility to exploit previous results.

\begin{figure}[t]
\centering
\includegraphics[width=0.9\columnwidth]{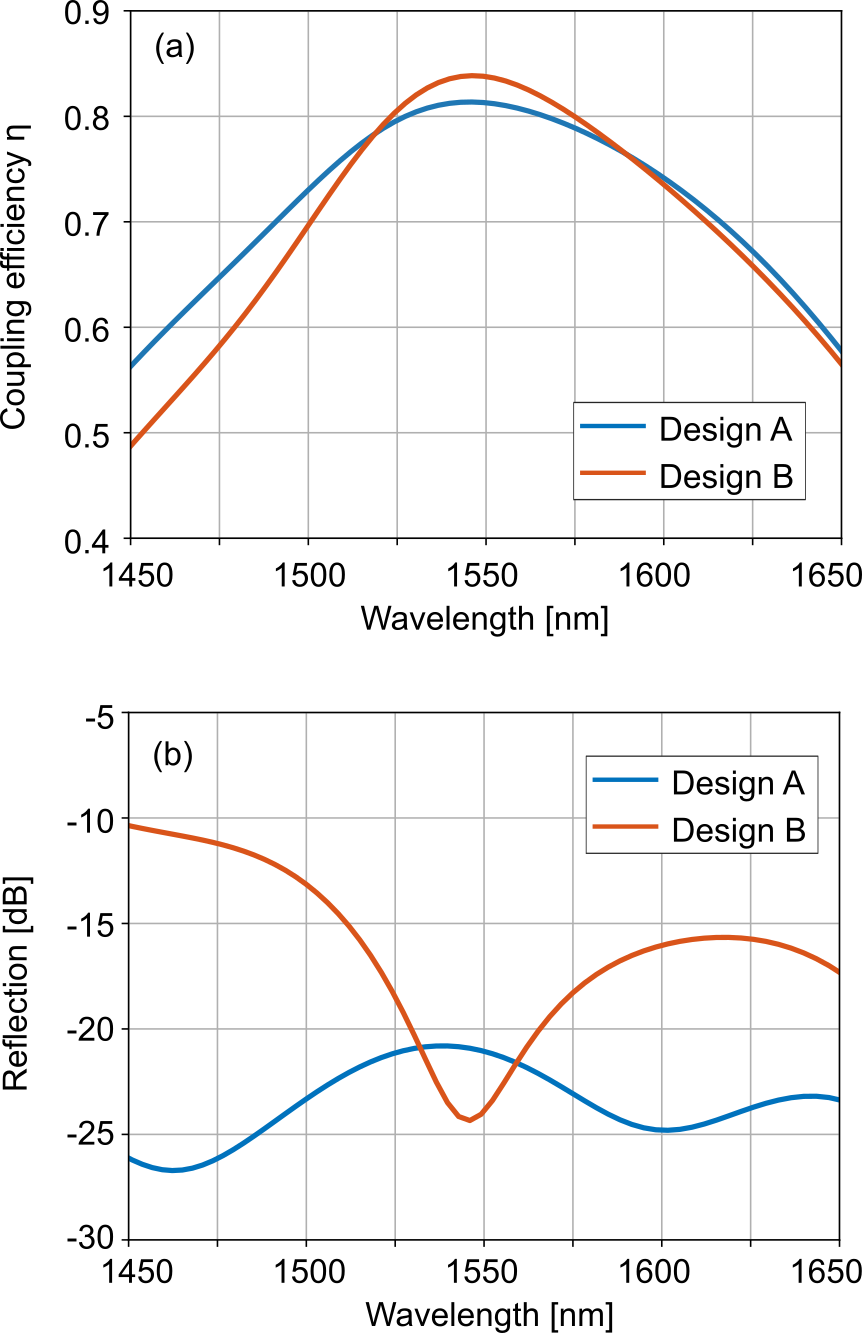}
\caption{(a) Coupling efficiency and (b) reflection spectra for the selected apodized design (blue solid lines) and for the apodized design with the highest coupling efficiency $\teta$ (orange solid line).}
\label{fig_apod_perforance}
\end{figure}

The structural parameters $[L_1^1, \cdots, L_5^2]$ of the design selected based on the requirements described in Eq. \ref{eq_opt_problem} (design A) are reported in Table \ref{table_designs}. The optimized apodized grating is very compact being only 3.6 \tmu{}m in length. As a reference, the parameters of the design that simply maximizes the coupling efficiency without any other criteria (design B) are also reported. The first period of the two gratings is rather different while the periodic section shows only minimal adjustments, mainly in the distance between the pillar and the L shape ($L_3^2$). The performance of these two designs are compared in Table \ref{table_apod_performance}. Figure \ref{fig_apod_perforance}(a) and (b) show the fiber coupling efficiency and reflection spectra, respectively, for the two designs. Design A and B have almost identical upward diffraction efficiencies $\rho_u$ of 0.919 and 0.910 at 1550 nm, respectively. This translates for design A in a fiber coupling efficiency $\teta$ of 0.813 (-0.9 dB), a small reduction compared to the maximum achievable $\teta$ of 0.838 (-0.77 dB, design B). A fiber misplacement of $\pm 0.8$ \textmu{}m from its ideal alignment with the antenna causes an additional coupling loss of about 1 dB. Directionality is very high and similar for both designs (0.973 and 0.978, respectively). The small penalty on efficiency for design A is compensated by a significant improvement of other performance metrics. The 1-dB bandwidth is about 30 nm larger than design B (157.9 nm instead of 128.9 nm). As a result, for design A $\teta$ only varies by 0.06 dB across the optical communication C band (from 1530 nm to 1565 nm) and 1.3 dB across the S, C, and L bands (from 1460 nm to 1625 nm). The reflection at 1550 nm is about 3 dB higher for design A, but over most of the considered wavelength range design B exhibits higher reflection, as shown in Fig. \ref{fig_apod_perforance}(b). Design A has a reflection consistently lower than -20 dB over the 200-nm wavelength range (R = -20.8 dB) while for design B reflection increases to -15 dB at $\lambda$ = 1510 nm and as high as -10.4 dB at $\lambda$ = 1450 nm.

\section{Fabrication tolerance}
\label{sec_tolerance}
\begin{figure*}[!ht]
\centering
\includegraphics[width=0.95\textwidth]{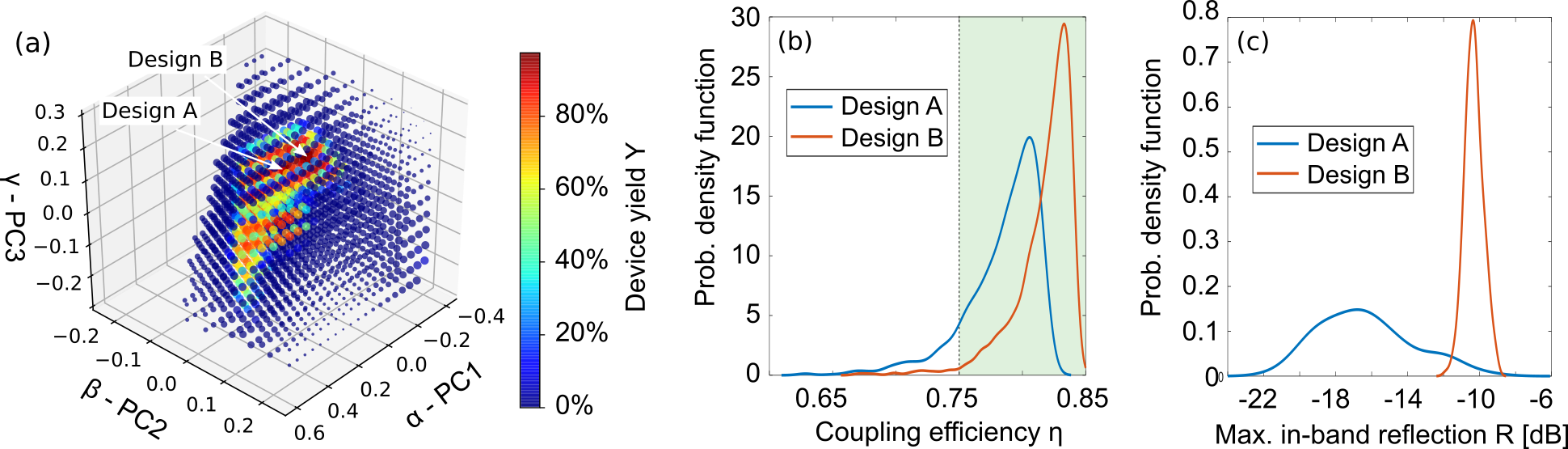}
\caption{Fabrication tolerance analysis for the one-step apodized grating design. (a) Device fabrication yield ($\teta \geq 0.75$) as a function of $\talpha$, $\tbeta$, and $\tgamma$. (b) Probability density function of the coupling efficiency and (c) the maximum reflection R in the 1450 nm - 1650 nm wavelength range for the selected design (deisgn A, solid blue curve) and the design with best $\teta$ (design B, orange solid line). The shaded green area in (b) highlights $\teta \geq 0.75$.}
\label{fig_tolerance}
\end{figure*}
The analysis of the impact of fabrication uncertainty on the antenna performance is a fundamental aspect of the design. Here we consider four different sources of common geometrical variability: Width deviations for the shallow and deeply etched sections, etch depth deviation, and a misalignment between the two etch steps. All these quantities are considered as normally-distributed independent random variables with zero mean and standard deviation of 5 nm for the width and etch variations and 10 nm for the misalignment. The stochastic analysis is performed very efficiently using a polynomial chaos model \cite{xiu_fast_2009, weng_stochastic_2017}. Twenty-five different values for the four variables are sampled according to their distribution. For a given nominal design, the corresponding twenty-five modified designs are generated and simulated to obtain the fiber coupling efficiency and the other performance metrics. A stochastic surrogate model to describe the dependence of $\teta$ on the four random variables (polynomial chaos model) is realized with second order Hermite polynomials as the orthonormal basis. The same is done for the maximum in-band reflection R. Finally, the probability density functions for $\teta$ and R are obtained with a standard Monte Carlo simulation by sampling the surrogate models $10^4$ times (which only takes a few seconds) and using a Gaussian kernel density estimator.

As a synthetic measure of the sensitivity of the fiber coupling efficiency to fabrication variability we compute the device fabrication yield Y as the probability for a given design to have $\teta \geq 0.75$. The yield is reported in Fig. \ref{fig_tolerance} (a) as a function of $\talpha$, $\tbeta$, and $\tgamma$. Comparing this figure with Fig. \ref{fig_apod_perforance} (b), it is evident how designs with similarly high fiber coupling efficiency do not necessarily tolerate fabrication variability in the same way, hence resulting in different yields. The region of the sub-space with high yield values is much smaller than the area with high fiber coupling efficiency in ideal conditions (i.e., no fabrication variability). Figure \ref{fig_tolerance} (b) shows in details the probability density function of $\teta$ for the design selected according to Eq. \ref{eq_opt_problem} (design A, blue solid curve) and the design with the highest fiber coupling efficiency (design B, orange solid line), as described in Sec. \ref{sec_optimization}. The two designs show a similar performance spread due to variability, with the latter slightly shifted towards higher values of $\teta$ because of its higher ideal efficiency. The green shade represents the condition $\teta \geq 0.75$, resulting, as reported in Tab. \ref{table_apod_performance}, in Y = 88\% and Y = 97\% for designs A and B, respectively. Despite a reduction of about 9\%, design A still shows a very high yield value which makes it a good candidate for fabrication. Lastly, Fig. \ref{fig_tolerance} (c) shows the probability density function for R (the highest reflection in the 1450 nm - 1650 nm wavelength range) for both designs. For design B (orange solid curve), R only undergoes minimal fluctuations due to fabrication variability but these fluctuations are centered at about -10 dB, the value for R in ideal conditions. Starting from this design, there is a negligible probability to obtain a fabricated antenna with $\mathrm{R} \leq -12$ dB. For design A (blue solid curve), R is much lower in ideal conditions (-20.8 dB) and not surprisingly this makes it more sensitive to fabrication variability. Despite this, there is a rather high probability of 64\% to obtain $\mathrm{R} \leq -15$ dB, confirming the good robustness of this design.

\section{Conclusion}
\label{sec_conclusions}
In this paper we have exploited a methodology based on adjoint optimization and machine learning dimensionality reduction for the multi-objective design optimization of a grating-based micro-antenna in a 300-nm SOI platform. The compact antenna is only 3.6 \textmu{}m long, has a perfectly vertical diffraction efficiency of almost 92\%, and directionality of 98\%. When coupled with an optical fiber with mode field diameter of 3.2 \textmu{}m vertically placed on top of the antenna, a coupling efficiency of more than 81\% is achieved with a wide 1-dB bandwidth of almost 158 nm. Reflection is smaller than -20 dB over the entire 1450 nm - 1650 nm wavelength range. These good performances make the antenna ideal for applications requiring dense arrays of both fiber and free-space coupling interfaces.


%

\appendices
\section{Design methodology details}
\label{app_1}
For the periodic grating, in the initial sparse set of good designs we only select optimized solutions with coupling efficiency $\teta \geq 0.6$ (17 designs), $\teta \geq 0.7$ (19 designs), and $\teta \geq 0.72$ (39 designs), for N = 3, N = 4, and N = 5, respectively. The maximum coupling efficiency among these designs is 0.612 (N = 3), 0.695 (N = 4), and 0.73 (N = 5), always smaller than the maximum efficiency found by sub-space mapping. The size of all these design sets is largely sufficient to guarantee the proper convergence of PCA \cite{melati_mapping_2019}. For N = 5, we consider twice the number of devices than the other cases to allow the analysis presented in Appendix \ref{app_2}. Two principal components allow to represent about 99\% of the variance for all the three sets, with a maximum projection error always smaller than 10 nm when measured in Manhattan distance $\sum_i|\widetilde{L_i}-L_i|$. In the three cases, we map the $\talpha$-$\tbeta$ sub-space by generating uniform grids of 2394, 925, and 1176 designs, respectively, at a Manhattan distance of 5 nm.
For the one-step apodized grating, we consider in the initial sparse set 16 good designs with $\teta \geq 0.75$ and a maximum coupling efficiency of 0.81. Three principal components are sufficient to represent 95\% of the set variance with a maximum projection error smaller than 16 nm. The map shown in Fig. \ref{fig_apod_design}(a) is generated with a grid of 2926 designs at a Manhattan distance of 100 nm while all other maps include 1502 designs at a distance of 50 nm.

\section{Physical insights on the grating behaviour}
\label{app_2}
\begin{figure}[!t]
\centering
\includegraphics[width=0.9\columnwidth]{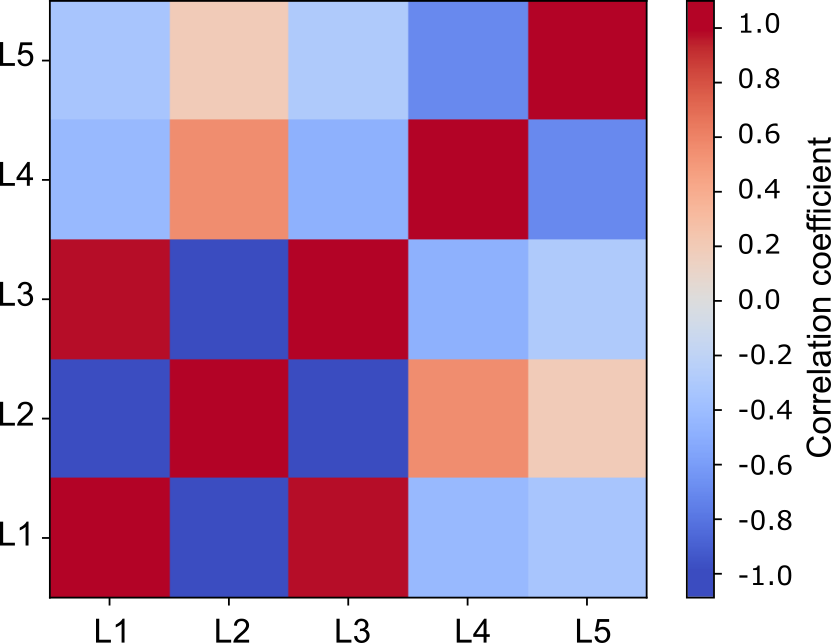}
\caption{Correlation matrix for the five geometrical parameters $[L_1, \cdots, L_5]$ of the periodic grating design with N = 5. The correlation coefficients are calculated using the initial sparse set of good designs.}
\label{fig_corr}
\end{figure}

Beside providing several advantages in the design, the use of dimensionality reduction makes it easier also in the case of multi-parameter devices to investigate how different parameters contribute to the device performance. To this purpose, the periodic grating design with N = 5 periods presented in Sec. \ref{sec_periodic} is considered here. For the five geometrical parameters $[L_1, \cdot, L_5]$, Fig. \ref{fig_corr} shows the correlation coefficient
\begin{equation}
\label{eq_rho}
    \rho_{i,j} = \frac{\mathrm{cov}(L_i,L_j)}{\sigma_{Li}\sigma_{Lj}}
\end{equation}
computed using the initial sparse set of 39 good designs (see Appendix \ref{app_1}). In Eq. \ref{eq_rho}, cov denotes the covariance and $\sigma$ is the standard deviation. The correlation coefficient measures the linear dependence between two parameters. As can be seen, the correlation matrix has a clear block structure. Parameter $L_1$ is strongly correlated with parameter $L_3$ ($\rho_{1,3} = 0.98$) and strongly anti-correlated with parameter $L_2$ ($\rho_{1,2}$ = -0.98) while correlation with the other two parameters is negligible. Likewise, parameters $L_4$ and $L_5$ are quite strongly anti-correlated ($\rho_{4,5} = -0.7$).

The application of PCA allows to highlight and take advantage of these correlations. The two principal components for this case (N = 5) are:
\begin{equation}
    \label{eq_periodic_V}
    \begin{split}
        \mathbf{V}_1 = [ 0.31, -0.18,  0.90, -0.21, -0.09], \\
        \mathbf{V}_2 = [-0.06, -0.02, -0.08, -0.70,  0.71].
    \end{split}
\end{equation}
For $\mathbf{V}_1$, the third component is at least three times larger than all of the other ones, meaning that $\mathbf{V}_1$ is largely aligned along the $L_3$ direction in the 5D design space. Because of the strong correlation between $L_1$, $L_2$, and $L_3$ we can conclude that $\mathbf{V}_1$ mostly represents the first three design parameters (which defines the silicon pillar and the two fully etched gaps, see Fig. \ref{fig_struct}(b)). This behaviour is even stronger for $\mathbf{V}_2$ where the fourth and fifth components are almost identical and about ten times larger than the other ones. We can hence assume $\mathbf{V}_2$ mostly represents $L_4$, and $L_5$ (the L-shaped structure). Because the goal of PCA is to represent a set of correlated parameters with a minimum number of independent parameters, the existence of two blocks of correlated parameters makes two principal components sufficient to represent the original 5D design space (or at least the portion of this space where good designs reside).

This first result can be used to analyse the behaviour of coupling efficiency (Fig. \ref{fig_ce_periodic}(c)) and reflection (Fig. \ref{fig_br_periodic}) as a function of $\talpha$ and $\tbeta$ (i.e., along $\mathbf{V}_1$ and $\mathbf{V}_2$ axes). Coupling efficiency is essentially circular in this 2D subspace, i.e., it depends in a similar way on both $\talpha$ and $\tbeta$. The same behaviour can be observed also for gratings with N = 3 (Fig. \ref{fig_ce_periodic}(a)) and N = 4 periods (Fig. \ref{fig_ce_periodic}(b)). This means that the entire grating period (both the initial pillar and the L-shaped structure) similarly contributes to coupling efficiency. On the other hand, reflection has a radically different dependence being mostly determined by the value of $\talpha$ with only limited variations along the $\mathbf{V}_2$ axis ($\tbeta$). It is hence mostly the initial pillar that allows to control reflection with limited contribution coming from the L shape. This analysis demonstrates how the use of dimensionality reduction could assist the designer intuition, giving access to a quantitative analysis of the behaviour of multi-parameter devices.



\ifCLASSOPTIONcaptionsoff
  \newpage
\fi



\bibliographystyle{IEEEtran}
\bibliography{IEEEabrv,micro_antenna_ref}
%



%
\newpage

\begin{IEEEbiography}[{\includegraphics[width=1in,height=1.25in,clip,keepaspectratio]{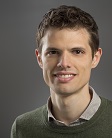}}]{Daniele Melati}
is a Research Associate at the national Research Council Canada. He received his Ph.D. in Information Engineering from Politecnico di Milano in 2014 working on device modelling for generic photonic foundries and distributed effects of waveguide roughness on guided light propagation. Currently, his research interests are related to novel applications of inverse design, optimization, stochastic techniques, and machine learning algorithms to efficiently design complex photonic devices. His research topics include also the design of ultra-efficient edge couplers and surface grating couplers based on subwavelength metamaterials.
\end{IEEEbiography}

\begin{IEEEbiography}[{\includegraphics[width=1in,height=1.25in,clip,keepaspectratio]{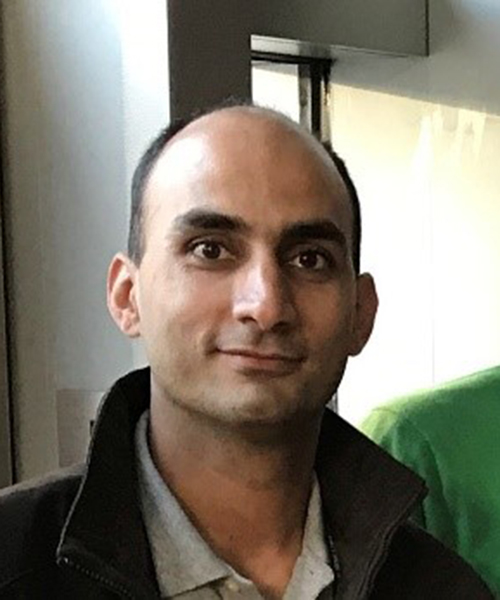}}]{Mohsen Kamandar Dezfouli}
received his Ph.D. degree in physics from Queen's University in Kingston, Ontario, Canada. He is currently a postdoctoral fellow with the Advanced Electronics and Photonics research center at National Research Council Canada. His current research interests are photonics, integrated photonics as well as the intersection between machine learning and optical design.
\end{IEEEbiography}

\begin{IEEEbiography}[{\includegraphics[width=1in,height=1.25in,clip,keepaspectratio]{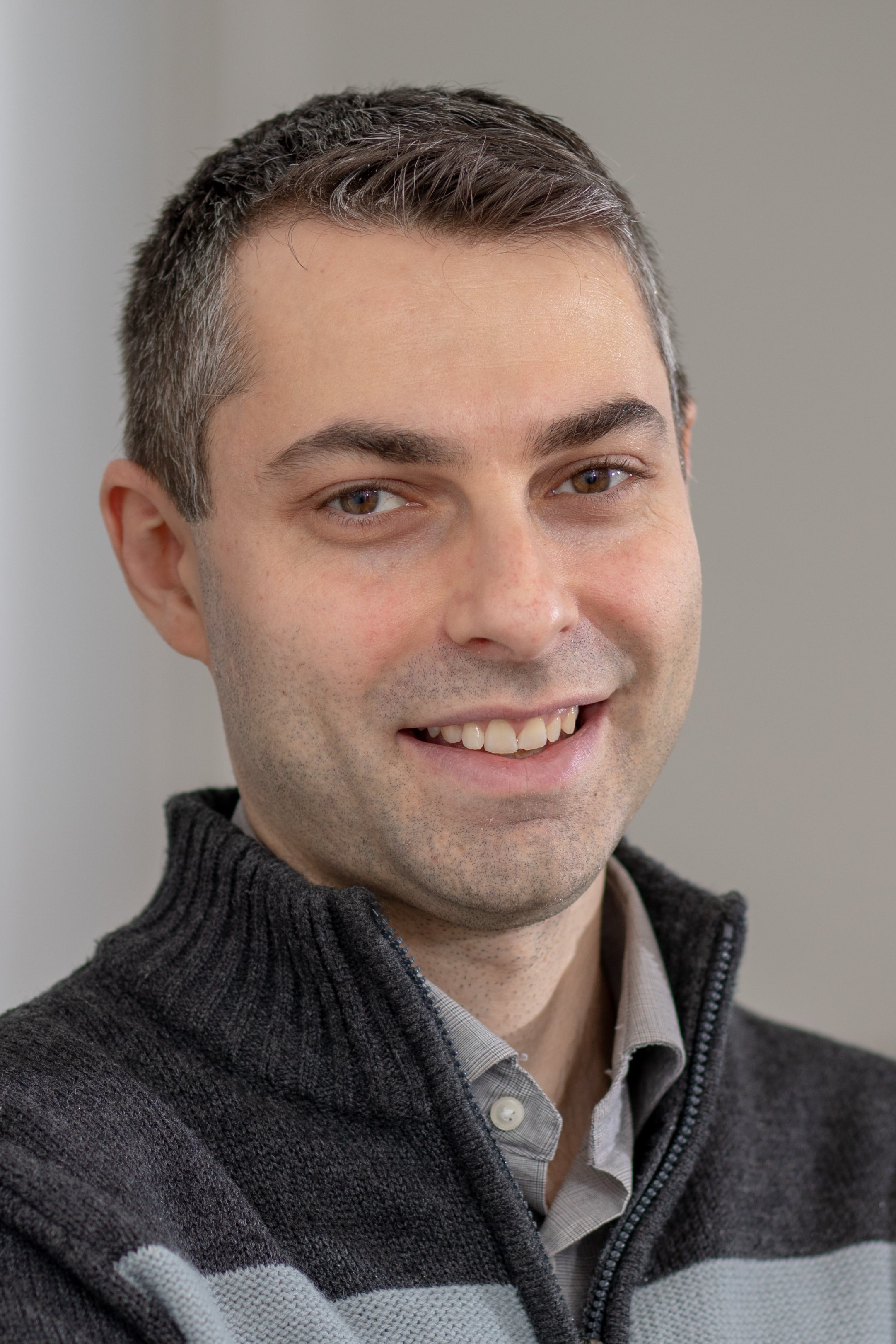}}]{Yuri Grinberg}
is an Associate Research Officer within the Digital Technologies research center, National Research Council of Canada. He obtained his PhD in Computer Science from McGill University at 2014 and was an NSERC postdoctoral fellow in Ottawa Hospital Research Institute. His expertise is applied and theoretical machine learning and reinforcement learning. He has co-authored over 20 peer-reviewed publications. In the past several years he has been working on the development of AI techniques for the design of nanophotonic components.
\end{IEEEbiography}

\begin{IEEEbiography}[{\includegraphics[width=1in,height=1.25in,clip,keepaspectratio]{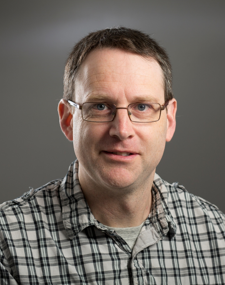}}]{Jens H. Schmid}
is a Senior Research Officer with the Advanced Electronics and Photonics Research Centre of the National Research Council Canada (NRC) and also an Adjunct Professor with the Department of Electronics at Carleton University. He received his Ph.D. degree from the University of British Columbia in 2004 for his work on in-situ etching and molecular beam epitaxial regrowth on III-V semiconductors. After working for a year as a research scientist for VSM MedTech, a medical device company in Coquitlam, B.C., where he developed fabrication processes for superconducting quantum interference devices, he joined the nanofabrication group at NRC in 2005. His current research interests are the fabrication, design, characterization and simulation of silicon photonic devices and nanostructures, in particular the use of subwavelength metamaterials in integrated photonic devices.  
\end{IEEEbiography}

\begin{IEEEbiography}[{\includegraphics[width=1in,height=1.25in,clip,keepaspectratio]{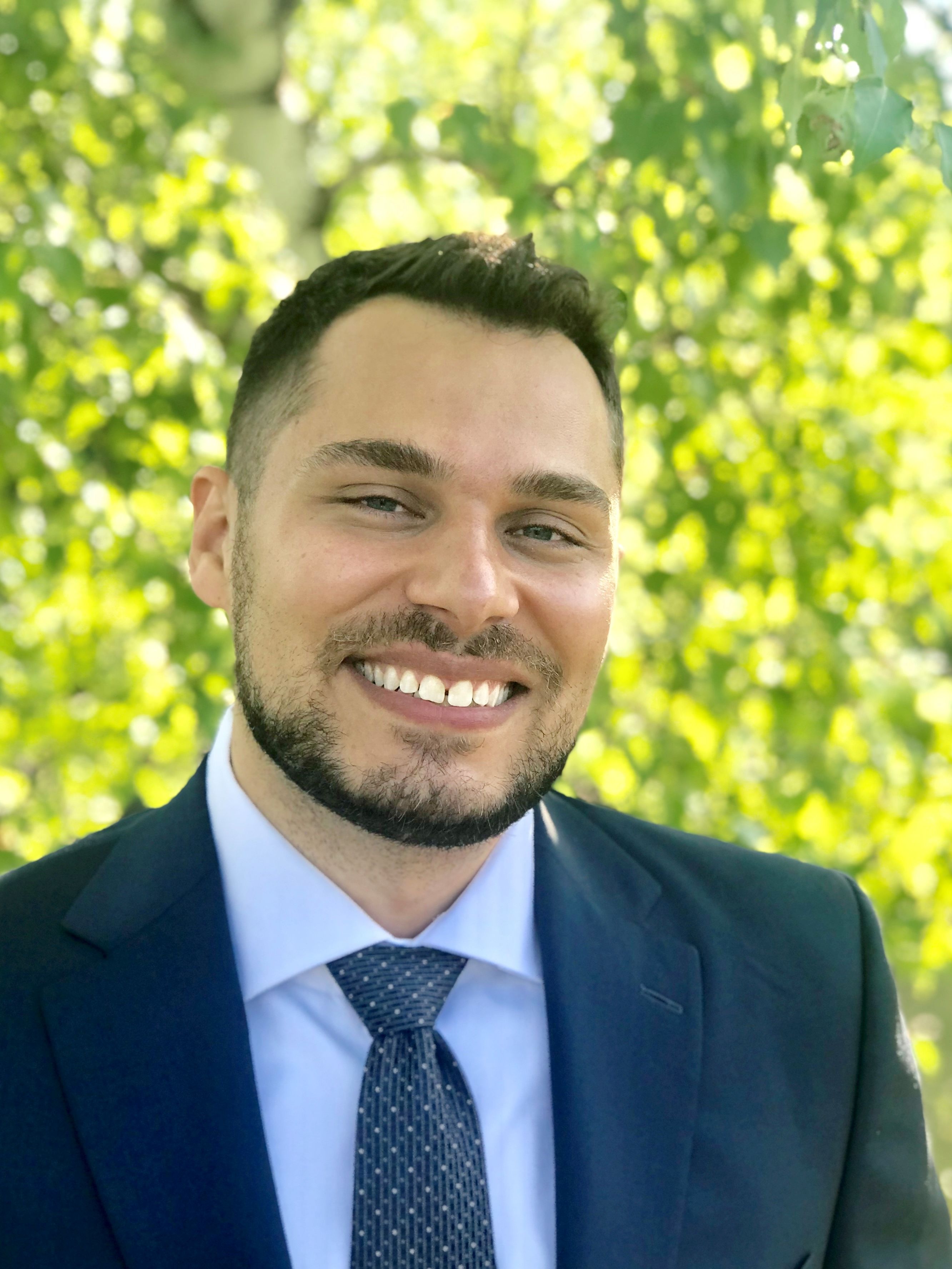}}]{Ross Cheriton}
is a Postdoctoral Fellow at the National Research Council Canada.  He received his PhD in Physics at the University of Ottawa’s SUNLAB on the theory and characterization of gallium nitride nanowires on silicon for intermediate band solar cells.  His research topics include integrated photonics, astrophotonics, III-V on silicon devices, optoelectronic simulations, gallium nitride optoelectronics, multijunction solar cells, nanowires, optical phototransducers,  quantum dot systems, entangled photon sources, applied control theory, optical systems for wireless epiretinal implants and free space data and power transmission and steering.  His work is currently focused on novel integrated astrophotonics devices for remote sensing and astronomy.
\end{IEEEbiography}

\begin{IEEEbiographynophoto}{Siegfried Janz}
has worked on silicon, glass, and III-V semiconductor integrated photonic devices at the National Research Council Canada (NRC) for more than 25 years, and was Program Leader for the Advanced Photonic Components Program at NRC from 2012 to 2018. His current research interests focus on silicon photonic devices for communications, astronomy, metrology and sensing. Dr. Janz completed his Ph.D. in physics in 1991 at the University of Toronto, working on nonlinear optics at metal surfaces. 
\end{IEEEbiographynophoto}

\begin{IEEEbiography}[{\includegraphics[width=1in,height=1.25in,clip,keepaspectratio]{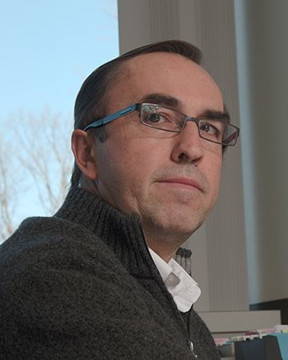}}]{Pavel Cheben}
is a Principal Research Officer at the National Research Council of Canada. He is also an official member of the Centre for Research in Photonics at the University of Ottawa, an Honorary Professor at University of Malaga and an Adjunct Professor at University of Toronto, Carleton University, University of Ottawa, McMaster University and University of Zilina. Dr. Cheben is best known for his work on subwavelength metamaterial waveguides, introducing a new area of science that brings together metamaterial research and integrated photonics. He has co-authored more than 550 papers and book chapters, 33 patent applications and over 250 invited presentations. He is a Fellow of the American Physical Society, the Optical Society of America, the European Optical Society, the Institute of Physics, the Engineering Institute of Canada, and the Canadian Academy of Engineering. He is a recipient of the Order of the Slovak Republic and of the Canada Public Service Excellence Award. He has been the most published scientist of the NRC Canada during the past ten years. 
\end{IEEEbiography}

\begin{IEEEbiography}[{\includegraphics[width=1in,height=1.25in,clip,keepaspectratio]{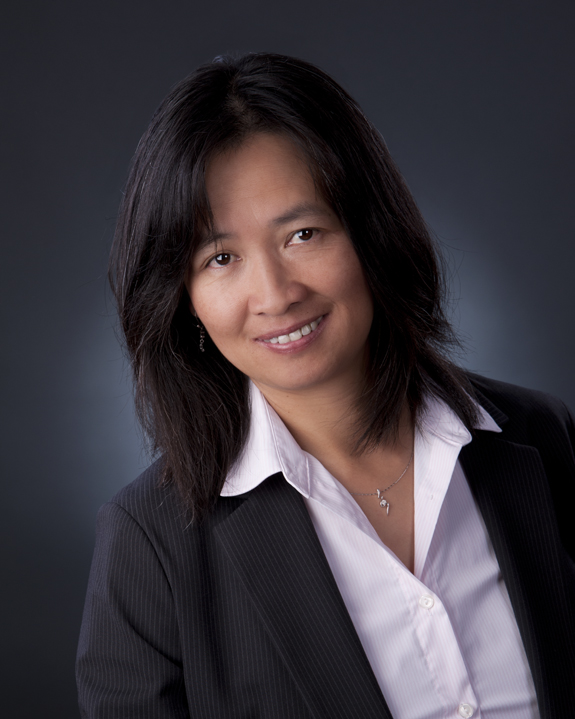}}]{Dan-Xia Xu}
is a Principal Research Officer with the National Research Council Canada, a Fellow of RSC and OSA, and an adjunct professor with Carleton University. Since joining NRC, her work encompassed high speed SiGe HBTs, silicides for sub-micron VLSI, SiGe and silicide photodetectors, and later integrated photonics. In 2001-2002 she was part of the team at Optenia Inc. that successfully developed the first commercial glass waveguide echelle grating demultiplexer. She has led the pioneering work in cladding stress engineering for polarization control of photonic components, and in high sensitivity biosensor systems using Si wire spiral resonators. Her current research interest is silicon photonics and nanophotonics for optical communications, sensing and thermometry. She particularly focuses on employing machine learning methods to guide the intelligent exploration of complex design space of nanophotonic components. She has co-authored over 400 publications, several book chapters, and holds seven patents.
\end{IEEEbiography}







\end{document}